\documentclass{iau} 
\usepackage{graphicx}

\title[Timescales after the AGB] 
{Evolutionary timescales \\ from the AGB to the CSPNe phase}

\author[Marcelo M. Miller Bertolami]   
{Marcelo M. Miller Bertolami$^{1,2}$}

\affiliation{$^1$Instituto de Astrof\'isica de La Plata, CCT-La Plata, CONICET-UNLP, \\ Paseo del Bosque s/n (B1900FWA), La Plata, Argentina \\ email: {\tt mmiller@fcaglp.unlp.edu.ar} \\[\affilskip]
$^2$Universidad Nacional de La Plata\\Paseo del Bosque s/n (B1900FWA), La Plata, Argentina}

\pubyear{2018}
\volume{xxx}  
\jname{Why galaxies care about AGB stars IV. IAUS 343.}
\editors{F. Kerschbaum, M. Gronewegen, H. Olofsson, \& V. Baumgartner, eds.}
\begin{document}

\maketitle

\begin{abstract}
The transition from the asymptotic giant branch (AGB) to the final
white dwarf (WD) stage is arguably the least understood phase in the
evolution of single low- and intermediate-mass stars ($0.8 \lesssim
M_{\rm ZAMS}/M_\odot\lesssim 8...10$). Here we briefly review the progress in
the last 50 years of the modeling of stars during the post-AGB
phase. We show that although the main features, like the extreme mass
dependency of post-AGB timescales were already present in the earliest
post-AGB models, the quantitative values of the computed post-AGB
timescales changed every time new physics was included in the modeling
of post-AGB stars and their progenitors. Then we discuss the
predictions and uncertainties of the latest available models regarding
the evolutionary timescales of post-AGB stars.

\keywords{stars: AGB
  and post-AGB, planetary nebulae: general, stars: evolution, stars:
  mass loss}
\end{abstract}

\firstsection 
\section{Modeling the evolution after the AGB. A historical introduction}
The transition from the asymptotic giant branch (AGB) to the final
white dwarf (WD) stage is arguably the least understood phase in the
evolution of single low- and intermediate-mass stars ($0.8 \lesssim
M_{\rm ZAMS}/\lesssim 8...10$). This transition phase includes the
so-called proto-planetary nebulae (PPNe) central stars and OH/IR
objects (also colectively known as \emph{post-AGB stars}, \cite[van
  Winckel 2003]{2003ARA&A..41..391V}\footnote{Note that throughout
  this paper we refer to the whole evolutionary stage between the end
  of the AGB and the beginning of the WD phase as the \emph{post-AGB
    phase}. This should not be confused with the so called
  \emph{post-AGB stars} which are defined as stars that have already
  departed from the AGB but are not yet hot enough to ionize the
  circumstellar material, i.e. $T_{\rm eff}\lesssim 30000$K.}) as well
as the hotter central stars of planetary nebula (CSPNe) and other
UV-bright stars.  In the most simple picture low- and
intermediate-mass stars undergo strong and slow stellar winds
($10^{-8}M_{\odot}$/yr$\lesssim\dot{M}\lesssim 10^{-4}M_{\odot}$/yr
and 3km/s$\lesssim v_{\rm wind}\lesssim 30$km/s, see \cite[H{\"o}fner
  \& Olofsson 2018]{2018A&ARv..26....1H}) at the end of the AGB which
lead to the almost complete removal of the H-rich envelope of the AGB
star ---see \cite{1957IAUS....3...83S}, \cite{1966PASP...78..232A} and
\cite{1970AcA....20...47P}.  After this point, the stars contract at
constant luminosity ($L_{\star}$) increasing their effective
temperatures ($T_{\rm eff}$) by almost two orders of magnitude, and if
there is enough material surrounding the star, a PN is formed in the process.

\begin{figure}[t]
\begin{center}
 \includegraphics[width=5.in]{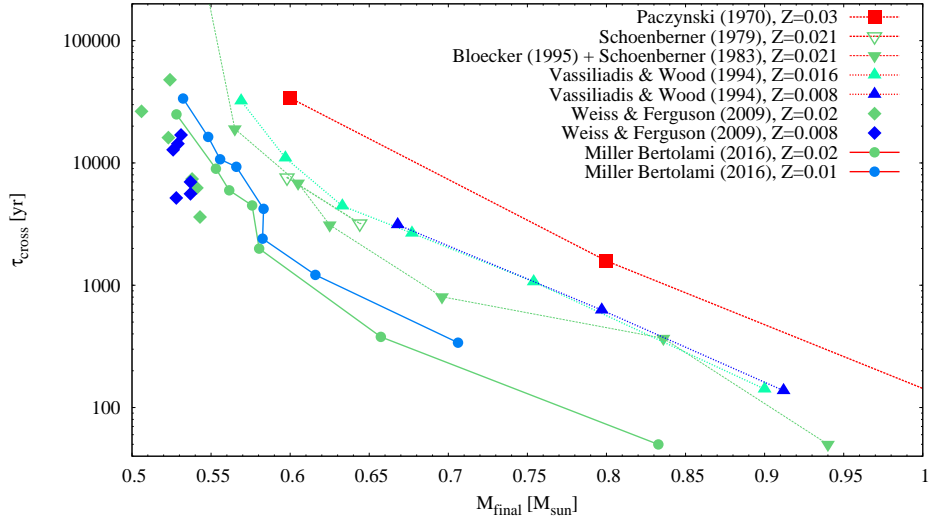} 
 \caption{Evolutionary post-AGB time  between the point at $\log T_{\rm eff} = 4$ to the point of maximum effective temperature in the HR diagram; $\tau_{\rm cross}$.}
   \label{fig1}
\end{center}
\end{figure}
Since the very first stellar evolution models of CSPNe were computed
by \cite{1970AcA....20...47P} it became clear that the evolution from
the AGB to the WD phase is extremely mass dependent. In fact, the
early models of \cite{1970AcA....20...47P} suggested that the time to
"cross'' the HR diagram decreased by 4 orders of magnitude just by
increasing the mass of the CSPN model by a factor of two (see
Fig. \ref{fig1}). The early models by \cite[ Paczy\'nski (1970,
  1971)]{1970AcA....20...47P, 1971AcA....21..417P} were constructed by
artificially fitting H-envelopes to core structures all obtained from
a flash suppressed AGB sequence of $M_i=3M_\odot$ ($Z=0.03$) and
evolved to obtain CO-cores at desired final remnant mass ($M_f=0.6,
0.8, 1.2M_\odot$).  \cite[Sch{\"o}nberner(1979,
  1981)]{1979A&A....79..108S, 1981A&A...103..119S} computed for the
fist time the transition from the AGB to the CSPNe phase by assuming a
steady wind according to the mass loss prescription by
\cite{1975psae.book..229R} and including a detailed computation of the
thermally pulsating (TP) AGB phase. This was done for two full
sequences with initial masses $M_i=1$ and $1.45 M_\odot$ (final masses
$M_f=0.598$ and $0.644 M_\odot$ respectively) and $Z=0.021$. These two
sequences already showed post-AGB timescales to be about 4.5 times
faster than predicted by the early \cite{1970AcA....20...47P} models
(see Fig. \ref{fig1}). Later, \cite{1983ApJ...272..708S} computed two
more sequences of initial masses $M_i=0.8$ and $1 M_\odot$ (initial
metallicity $Z=0.021$, final masses $M_f=0.546$ and $0.565 M_\odot$
respectively) by including for the first time a "superwind'' phase at
the end of the AGB with mass-loss rates of $\dot{M}\gtrsim 10^{-4}
M_\odot$/yr. Although it only covered a small mass range,
Sch{\"o}nberner's post-AGB models were the first to include a detailed
treatment of the TP-AGB, showing the importance of AGB modeling for
the computation of accurate post-AGB stellar models
\cite[(Sch{\"o}nberner 1987)]{{1987fbs..conf..201S}}. The next grid of models, which covered a
wider range of remnant masses ($0.6\leq M_{f}/M_\odot\leq 0.89$), was
computed by \cite{1986ApJ...307..659W}. These models were constructed
by artificially stripping most of the H-envelope from red giant models
computed through many thermal pulses on the AGB but from a single
progenitor sequence of $M_i=2 M_\odot$ ($Z=0.02$).
\cite{1986ApJ...307..659W} computed the end of the TP-AGB by assuming
two different extreme mass loss rates in their computations. However,
the lack of a realistic initial-final mass relation \cite[(IFMR;
  Weidemann 1987)]{1987A&A...188...74W} had consequences in the
predicted post-AGB evolution and was criticized by
\cite{1990A&A...240L..11B}.

 The following generation of post-AGB models came in the 90's when
 both \cite{1994ApJS...92..125V} and \cite{1995A&A...299..755B}
 published grids of post-AGB models, covering the whole mass range of
 CSPNe, which included a detailed treatment of the winds during the
 TP-AGB phase ---see \cite{1993ApJ...413..641V} and
 \cite{1995A&A...297..727B}. In particular these grids adopted
 different initial masses to produce different CSPNe, as expected from
 early determinations of the Initial/Final Mass Relation (see
 \cite[Weidemann 1987]{1987A&A...188...74W} and references
 therein). These grids of post-AGB stellar evolution models
 represented a great improvement over the previous \cite[ Paczy\'nski
   (1970, 1971)]{1970AcA....20...47P, 1971AcA....21..417P} and
 \cite{1986ApJ...307..659W} models and confirmed the previous result
 by \cite[Sch{\"o}nberner(1979, 1981)]{1979A&A....79..108S,
   1981A&A...103..119S} that post-AGB timescales were several times
 shorter than predicted by \cite{1970AcA....20...47P}, as can be seen
 in Fig. \ref{fig1}. It is worth noting, however, that neither
 \cite{1994ApJS...92..125V} nor \cite{1995A&A...299..755B}
 incorporated the impact of core-overshooting in the upper main
 sequence, which was already know at the time to be significant, e.g.
 \cite{1992A&AS...96..269S}. Neither did these models make use of the
 updated radiative opacities computed by the OPAL \cite[(Rogers \&
   Iglesias 1992)]{1992ApJS...79..507R} and OP \cite[(Seaton et
   al. 1994)]{1994MNRAS.266..805S} projects that revolutionized
 stellar astrophysics during the early nineties.

\begin{figure}[t]
\begin{center}
 \includegraphics[width=4.5in]{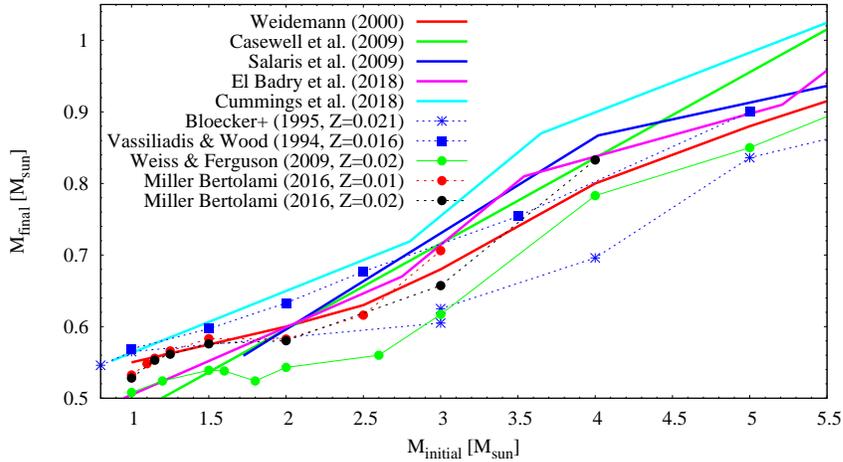} 
 \caption{Initial-Final Mass Relation (IFMR) of different post-AGB
   stellar evolution models as compared with the latest semiempirical
   determinations \cite[(Casewell et al. 2009, Salaris et al. 2009, El
     Badry et al. 2018 and Cummings et al. 2018)]{2009MNRAS.395.1795C,
     2009ApJ...692.1013S, 2018ApJ...860L..17E, 2018arXiv180901673C},
   and the classic semiempirical relation of
   \cite{2000A&A...363..647W}.}
   \label{fig2}
\end{center}
\end{figure}
 About a decade later another significant improvement in AGB and
 post-AGB stellar evolution models was made by
 \cite{2008PhDT.......290K} (later published in \cite[Weiss \&
   Ferguson 2009]{2009A&A...508.1343W}, from now on KWF). These
 authors incorporated several features for the first time, both in
 the computation of the AGB and post-AGB stellar evolution
 models. First, following \cite{2002A&A...387..507M}, these authors
 included for the first time C-rich molecular opacities in the
 computation of full AGB stellar evolution models. In addition, they
 also incorporated a separated treatment of C-rich and O-rich
 dust-driven winds. Most importantly, these authors included both the
 impact of convective boundary mixing on the main sequence, helium
 core-burning stage and TP-AGB evolutionary stages, as well as the
 inclusion radiative opacities from the OPAL project. Probably because
 of the latter, many convergence problems prevented KWF from computing
 a large grid of post-AGB stellar evolutionary models. In spite of the
 lack of a large grid of post-AGB sequences, the models computed by
 KWF already showed a clear trend, the post-AGB evolution of these
 models was much faster than those computed by
 \cite{1994ApJS...92..125V} or \cite{1995A&A...299..755B}, see
 Fig. 1. Again, as it happened with Sch{\"o}nberner's post-AGB
 models more than two decades before, an improvement in the modeling
 of previous evolutionary stages lead to much shorter post-AGB
 timescales.  Finally, following the approach of KWF, \cite[Miller
   Bertolami (2015, 2016)]{2015ASPC..493...83M, 2016A&A...588A..25M}
 computed a larger grid of post-AGB stellar evolution models. The main
 difference between this work and that of KWF is that mixing at
 convective boundaries from the ZAMS to the TP-AGB were calibrated
 trying to reproduce several observables on the main sequence and on
 the TP-AGB and post-AGB evolutionary stages. In particular, the
 models computed by Miller Bertolami are able to reproduce the width
 of the main sequence, the C/O ratios of the AGB and post-AGB stars
 and the He, C and O abundances observed in PG1159 stars. Most
 importantly, the IFMR of the theoretical models computed by
 \cite{2016A&A...588A..25M} are closer to the semiempirically derived
 ones than those of KWF (see Fig. \ref{fig2}).  In agreement with the
 findings of KWF the post-AGB models computed by
 \cite{2016A&A...588A..25M} are significantly faster and slightly
 brighter than earlier models of similar final mass (see
 Fig. \ref{fig1}).




\section{Post-AGB timescales and definitions}
 The terminology used to define the various stages after the departure
 from the AGB is sometimes confusing.  For example, stellar evolution
 studies usually refer to the whole stage between the end of the AGB
 and the beginning of the white dwarf phase as the post-AGB stage
 (e.g. \cite[Vassiliadis \& Wood 1994, Bl{\"o}cker 1995 and Miller
   Bertolami 2016]{1994ApJS...92..125V, 1995A&A...299..755B,
   2016A&A...588A..25M}), while observationally is common to refer as
 \emph{post-AGB stars} to those stars that have already departed from
 the AGB but are not yet hot enough to ionize the circumstellar
 material, i.e. $T_{\rm eff}\lesssim 30000$K  \cite[(van Winckel
   2003, Szczerba et al. 2007)]{2003ARA&A..41..391V,
   2007A&A...469..799S}. Also, from the observational point of view it
 is usual to split the evolution after the AGB and before the WD phase
 into the Proto-Planetary Nebulae (PPNe) and the Central Star of
 Planetary Nebulae (CSPNe) phases.  Within this classification the
 PPNe phase corresponds to the early evolution from the end of the AGB
 to the beginning of the ionization of the surrounding nebulae at
 about $T_{\rm eff}\sim 30000$K. During the PPNe phase it is expected
 that many central stars would still be enshrouded in dust and not
 visible in the optical, see \cite{2007A&A...469..799S}. The CSPNe
 phase would then correspond to the phase from the moment the central
 star attains $T_{\rm eff}\sim 30000$K to the beginning of the white
 dwarf cooling track. Yet, from the point of view of the evolution of
 the central star, this classification is of little use, as it relies
 on the properties of the surrounding material. Even more, very low
 mass stars might not evolve fast enough or eject significant amounts
 of material during the late AGB phase to produce a visible PNe.

\begin{figure}[t]
\begin{center}
 \includegraphics[width=3.in, angle=-90]{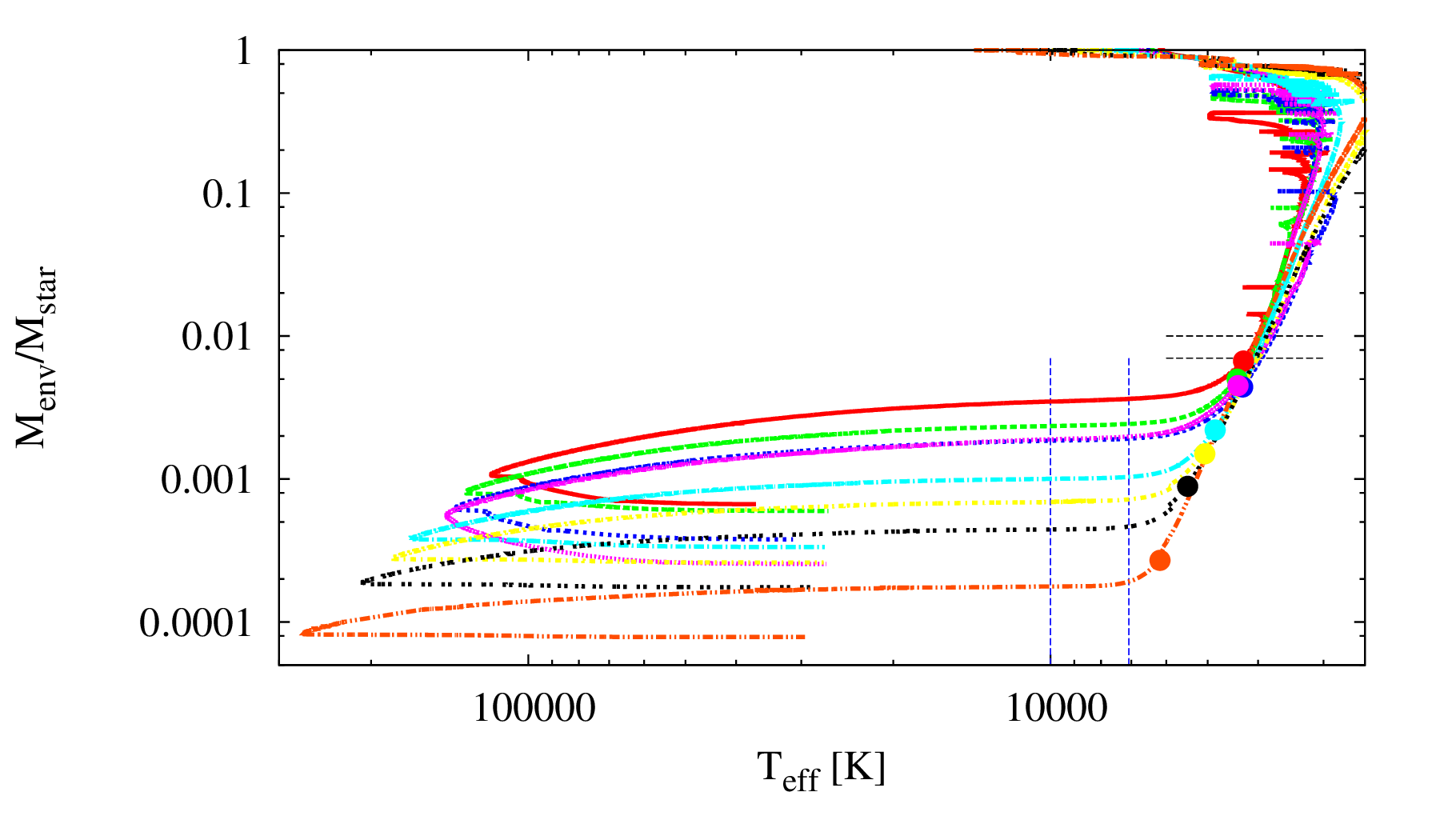} 
 \caption{Mass of the H-rich envelope of the H-burning models computed
   by \cite{2016A&A...588A..25M}. The vertical dashed lines indicate the zero points at $\log T_{\rm eff}=3.85$ ($\log T_{\rm eff}=4$) adopted by \cite{2016A&A...588A..25M} \cite[(Vassiliadis \& Wood 1994)]{1994ApJS...92..125V} for the computation of the post-AGB crossing times ($\tau_{\rm cross}$). Horizontal dashed lines indicate two alternative envelope masses adopted by \cite{2016A&A...588A..25M} as a definition of the end of the AGB. Color circles indicate the end of the AGB as estimated from the suggestion by \cite{2008ApJ...674L..49S}.}
   \label{fig3}
\end{center}
\end{figure}
 In order to be able to quantify the properties of the models during
 the post-AGB phase precise definitions are required.  In
 particular it is worth noting that the very idea of the end of the
 AGB is not easy to define from the point of view of stellar
 evolution, as stars continuously evolve away from the AGB during the
 late AGB evolution but without any sudden change in the stellar
 properties From the point of view of the structure of the central
 star the main change that takes place during the departure from the
 AGB is the transition from a expanded giant-like configuration into a
 dwarf-like one. This is caused by the reduction of the H-rich
 envelope below the critical value required to sustain a giant-like
 structure (see Fig \ref{fig3} and \cite[Faulkner
   2005]{2005slfh.book..149F} for an extended discussion of this
 problem). 
\begin{figure}[t]
\begin{center}
 \includegraphics[width=3.in, angle=-90]{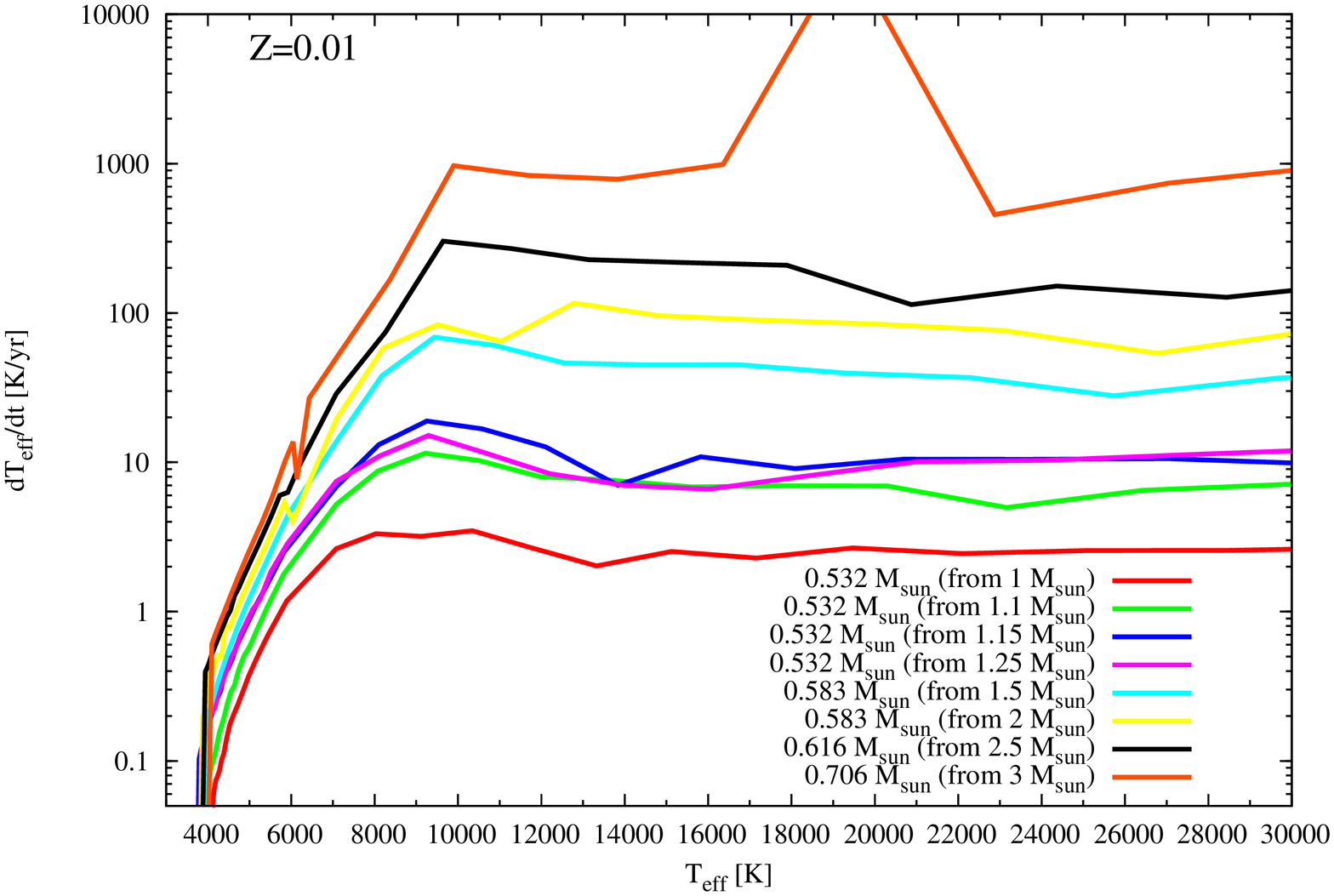} 
 \caption{Heating speed of the stellar evolution models during the
   early post-AGB phase. Note the exponential increase in the heating
   rate below $T_{\rm eff}\lesssim 10000$K as the almost constant
   heating rate at $10000{\rm K}\lesssim T_{\rm eff}\lesssim 30000{\rm
     K}$. }
   \label{fig5}
\end{center}
\end{figure}
This leads to a continuous increase in the heating rate of the stellar
surface from $\dot{T}_{\rm eff}\lesssim 0.1$ K/yr on the AGB to $1
  {\rm K/yr}\lesssim \dot{T}_{\rm eff}\lesssim 10000 {\rm K/yr}$ once
    the star attained $T_{\rm eff}\gtrsim 10000 {\rm K}$, see
      Fig. \ref{fig5}.

In this context, and in order to discuss the properties of the
computed stellar models, different authors choose to divide the
transition from the AGB to the WD phase according to different
arbitrary definitions. As the relative mass of the envelope is a key
feature determining the end of the AGB, \cite{2016A&A...588A..25M}
choose to define the end of the AGB
phase as the moment in which $M_{\rm env} /M_\star = 0.01$ (see Fig
\ref{fig3}). At this moment, models have already moved significantly
to the blue ($T_{\rm eff}\sim 3700... 5000$K), which is true at all masses and metallicities. Although
this choice defines the end of the AGB in a homogeneous way for all
masses and metallicities, and is based on the underlying physical
reason behind the departure from the AGB, the choice remains arbitrary.

In order to disentangle the impact of different uncertainties and
definitions we define two different timescales: the transition time
scale $\tau_{\rm trans}$ corresponding to the early (and slow) evolution
from the end of the AGB ($M_{\rm env} /M_\star = 0.01$) to the point
at $\log T_{\rm eff}=3.85$, and the crossing timescale $\tau_{\rm
  cross}$ corresponding to the late (fast) evolution from $\log T_{\rm
  eff}=3.85$ to the point of maximum effective temperature.
In what follows we discuss the properties and uncertainties of these two post-AGB timescales.
\begin{figure}[t]
\begin{center}
 \includegraphics[width=3.in, angle=-90]{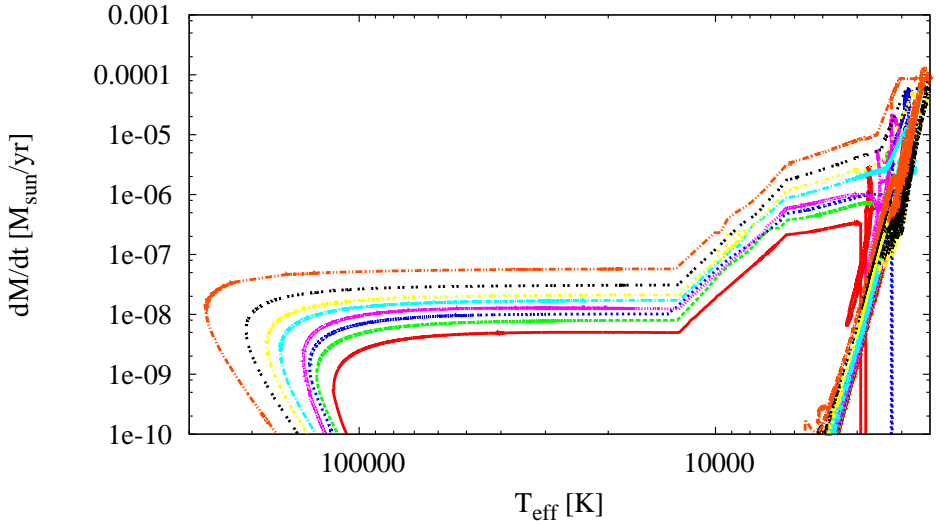} 
 \caption{Mass loss rates of the sequences computed by
   \cite{2016A&A...588A..25M} for $Z=0.01$.}
   \label{fig4}
\end{center}
\end{figure}

\subsection{The crossing time: $\tau_{\rm cross}$}
The uncertainties in $\tau_{\rm cross}$ (Fig. \ref{fig1}) are mostly
related to uncertainties in the previous evolution, and not to
uncertainties of the physics during the post-AGB phase itself. In
particular, with the exception of very luminous CSPNe one should not
expect hot radiative-driven winds to be of any importance for the
value of $\tau_{\rm cross}$ in H-burning post-AGB stars.
Fig. \ref{fig4} shows the evolution of the mass loss rate $\dot{M^{\rm
    env}}_{winds}$ adopted in the computation of the sequences
computed by \cite{2016A&A...588A..25M} for $Z=0.01$.  The rate of reduction 
of the H-rich envelope by winds have to be compared with the rate of
H-consumption by nuclear burning which in the case of CNO-burning is
of $\dot{M_H}/(M_\odot$yr$^{-1})\sim 10^{-11}
L_H/L_\odot$. Consequently, for typical post-AGB stars, in the range
$\log L/L_\odot=3...4$, the H-rich envelope is consumed by nuclear
burning at $\dot{M}^{burn}_{\rm env}\sim 1.4\times 10^{-8} ... 2\times
10^{-7}$. Consequently, and with the exception of the more massive and
luminous model, winds affect the rate of consumption of the H-rich
envelope by less than a 20\%, see Table 3 of
\cite{2016A&A...588A..25M}, and do not play a significant role in the
determination of $\tau_{\rm cross}$. The value of $\tau_{\rm cross}$
is consequently determined by the mass of the H-rich envelope at which
the model departs from the AGB and the intensity of the H-burning
shell. While these two properties are to some extent affected by the
phase of the thermal pulse cycle at which the star departs from the
TP-AGB, they are much more affected by the degeneracy level of the
CO-core and intershell \cite[(see Bl{\"o}cker
  1995)]{1995A&A...299..755B} as well as by the chemical composition
of the H-rich envelope \cite[(see Tuchman et al. 1983 and Marigo et
  al. 1999)]{1983ApJ...268..356T, 1999A&A...351..161M}. In turn these
properties are mostly affected by the microphysics adopted in the
models and the macrophysics (winds and convective boundary mixing
prescriptions) which affect the efficiency of third dredge up as well
as the length of the TP-AGB phase and the initial-final mass relation.
It is worth emphasizing that convective boundary mixing the main
sequence also affects the final post-AGB timescales, as it has an
important impact in the initial final mass relation, see
\cite{2009ApJ...692.1013S} and also Wagstaff et al. in preparation.

Additionally, due to the fast evolution from $T_{\rm eff}=$7000 K to
10000K (see Fig. \ref{fig5}) different choices of the zero point (like
those of \cite[Vassiliadis \& Wood 1994, Bl{\"o}cker 1995 and Miller
  Bertolami 2016]{1994ApJS...92..125V, 1995A&A...299..755B,
  2016A&A...588A..25M}) have a negligible impact in the actual value
of $\tau_{\rm cross}$. This, together with the previously discussed
points,  link all important
uncertainties and differences in the computations of $\tau_{\rm
  cross}$ (see Fig. \ref{fig1}) directly to the modeling of previous
evolutionary stages.

\subsection{The transition time: $\tau_{\rm trans}$}
At variance with what happens with $\tau_{\rm cross}$ the value of the
transition time $\tau_{\rm trans}$ is directly affected by the
intensity of stellar winds during this phase. As can be seen in
Fig. \ref{fig4} mass loss rates are well above the threshold of
$10^{-8}...10^{-7} M_\odot/$yr and the evolution is then dominated by
the intensity of stellar winds. To make things worst, winds during
this transition phase are completely uncertain. Also, attempts to
measure the evolutionary speed of these objects by means of the study
of the period drift in PPNe are not still possible (see Hrnivak et
al., these proceedings). For example, \cite{2016A&A...588A..25M}
adopted during this stage mostly the wind prescription for cold giant
winds by \cite{2005ApJ...630L..73S} which have only been validated for much cooler stars
$T_{\rm eff}\lesssim 4500{\rm K}$\cite[(see Schr{\"o}der \& Cuntz 2007)]{2007A&A...465..593S}, so inaccuracies of a factor of a
few are not unthinkable. Note that, as $\tau_{\rm trans}$ is basically
determined by speed of the reduction of the remaining H-rich envelope
by winds, any error in the wind intensity in this transition regime
($4000{\rm K}\lesssim T_{\rm eff}\lesssim 7000{\rm K}$) will directly
translate into errors in the computed value of $\tau_{\rm trans}$.

In addition to our current lack of knowledge of winds during this
early post-AGB phase the arbitrariness in the definition of the end of
the AGB (an thus of the beginning of this early stage) directly
affects the value of $\tau_{\rm trans}$.
\begin{figure}[t]
\begin{center}
 \includegraphics[width=4.5in]{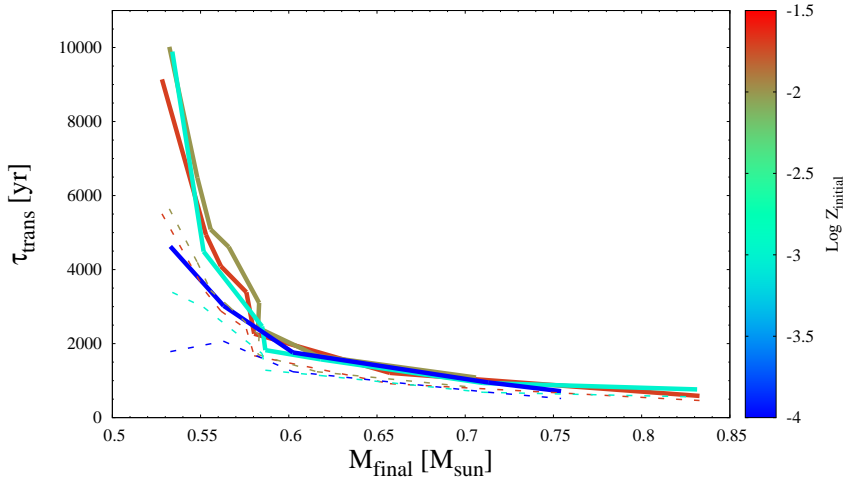} 
 \caption{Solid lines: Transition times $\tau_{\rm trans}$ from the
   end of the AGB defined at $M_{env}=0.01 M_\star$ to the point at
   $\log T_{\rm eff}=3.85$ during the post-AGB evolution for sequences
   of different final masses and metallicities \cite[(from Miller
     Bertolami 2016)]{2016A&A...588A..25M}. Dashed lines: Same as the
   solid lines but with the end of the AGB defined as the point at
   $M_{env}=0.007 M_\star$. }
   \label{fig5}
\end{center}
\end{figure}
Fig. \ref{fig5} shows that the value of $\tau_{\rm trans}$ would
change by a factor of $\sim 2$ if the end of the AGB would have been
defined as $M_{env}=0.007 M_\star$ (dashed lines) as compared with the
$M_{env}=0.01 M_\star$ adopted by \cite{2016A&A...588A..25M}. In this
connection it is worth recalling the suggestion by
\cite{2008ApJ...674L..49S} of quantitative definition for the end of
the TP-AGB based on the ratio $Q$ of the dynamical and envelope
timescales of the star. Fig. \ref{fig3} shows a simple estimation of
the point at which $Q$ reaches its maximum value (from Eq. 6 in
\cite[Soker 2008]{2008ApJ...674L..49S} and under the assumption of
$\beta=1$). Fig. \ref{fig3} suggests that the criterion proposed by
\cite{2008ApJ...674L..49S} might indeed be able to capture key aspect
of the transition from the AGB to the post-AGB, as it defines the end
of the AGB close to the point where the fast post-AGB phase starts. It
remains to be seen to which extent it agrees with the observational
definitions of the post-AGB phase, but it certainly deserves further
examination.

In view of the previous discussion, the values of $\tau_{\rm trans}$
are only qualitatively useful. In particular, an interesting result
from Fig. \ref{fig5} is that $\tau_{\rm trans}$ changes by more than
an order of magnitude when going from $M_f\sim 0.5 M_\odot$ to
$M_f\gtrsim 0.7 M_\odot$. Note, in particular, that for $M_f\gtrsim
0.6 M_\odot$ this stage lasts for $\tau_{\rm trans}\lesssim 2000$ yr
for all metallicities.

\section{Final comments}
During the last 50 years our modeling of post-AGB stars has been
slowly but continuously improved as better physics (both micro- and
macro-physics) have been included in the modeling both of the post-AGB
and previous evolutionary phases. In particular, it seems that with
each current improvement post-AGB timescales became shorter. Current
models that have been calibrated to reproduce several observables in
the evolution of low- and intermediate-mass stars \cite[(Miller
  Bertolami 2016)]{2016A&A...588A..25M} indicate that the time
required to cross the HR-diagram from $T_{\rm eff}\sim 7000$K to
$T_{\rm eff}\gtrsim 7000$K is of only $\tau_{\rm cross}\sim 10000$yr
for remnant stars of $M_f\sim 0.55 M_\odot$ of $\tau_{\rm
  cross}\lesssim 2000$yr $M_f\gtrsim 0.58 M_\odot$ and less than a few
hundred years for objects with $M_f\gtrsim 0.70 M_\odot$. The fast
post-AGB evolution of this models helps to explain the observed
existence of single CSPNe of low mass \cite[(e.g. Althaus et al.2008,
  Henry et al.2018 and Miller et al.2018)], as well as to understand
the properties of CSPNe in the Galactic Bulge \cite[(Gesicki et
  al. 2014)]{2014A&A...566A..48G}. In addition the faster evolution of
current post-AGB models might be key to understand the mystery of the
invariance of the planetary nebulae luminosity cut-off mystery
\cite[(Gesicki et al.2018)]{2018NatAs...2..580G} and, may be, the
dearth of post-AGB stars in M32 \cite[(Brown et
  al.2008)]{2008ApJ...682..319B}. However, although the current models
are able to reproduce several observables of AGB and post-AGB
populations \cite[(Miller Bertolami 2016)]{2016A&A...588A..25M}, some
significant discrepancies are still present. The most important ones
are the inability of the current models to reproduce the total
lifetime of intermediate luminosity M-stars and C-stars at about the
LMC luminosity (see \cite[Miller Bertolami 2016]{2016A&A...588A..25M})
and the systematically lower final masses of current models in the
range $M_i\sim 2...3 M_\odot$ when compared with the latest
determinations of the initial-final mass relation, see Fig. \ref{fig2}
\cite[(Casewell et al. 2009, Salaris et al. 2009, El Badry et al. 2018
  and Cummings et al. 2018)]{2009MNRAS.395.1795C, 2009ApJ...692.1013S,
  2018ApJ...860L..17E, 2018arXiv180901673C}. A lower intensity of
third dredge up processes and a lower intensity of the mass loss
during the C-rich phase might help to solve both problems (Wagstaff et
al., in preparation).

Still, the largest uncertainty in our current understanding of the
post-AGB evolution in single stars concerns the intensity of winds
during the departure from the AGB $4000 {\rm K}\lesssim T_{\rm
  eff}\lesssim 7000 {\rm K}$ which strongly affects the evolutionary
speed of the models during the transition stage ($\tau_{\rm trans}$).

 Finally it should be mentioned that all post-AGB stellar
evolutionary sequences are based on the assumption that the final
ejection of the envelope occurs through steady winds. This leads to a
well defined relationship between the mass of the remnant and the mass
of the remaining H-rich envelope. The strong dependency of the critical
mass of the envelope at which the stars depart from the AGB
(Fig. \ref{fig3}) is key in the determination of the mass dependency of
the crossing timescale ($\tau_{\rm cross}$, Fig. \ref{fig1}). If some
objects eject their envelopes by means of a dynamical phase due to
binary interaction or, for example, the ingestion of a substellar
companion then the remnant might depart from the AGB with smaller
envelope masses and our of thermal equilibrium \cite[(e.g. Hall et
  al.2013)]{2013MNRAS.435.2048H}, and evolve through the post-AGB
phase at a much faster pace. In particular, this implies that any
comparison of post-AGB stellar models like those computed by
\cite[Vassiliadis \& Wood (1994), Bl{\"o}cker (1995) or Miller
  Bertolami (2016)]{1994ApJS...92..125V, 1995A&A...299..755B,
  2016A&A...588A..25M}) with CSPNe in close binary systems should be
address with strong skepticism.

\acknowledgements M3B thanks the IAU and the organizers for a travel
grant and the waiving of the registration fee which allowed him to
attend this exciting Symposium. M3B also acknowledges a travel grant
from the Facultad de Ciencias Astron\'omicas y Geof\'isicas.

\begin{discussion}

\discuss{D'Antona}{From your very complete presentation, do I understand correctly that, in spite of the difficulties with the determination of the transition time, massive planetary nebulae nuclei cannot be found in the high luminosity crossing phase, like in the old Paczynski models?}

\discuss{Miller Bertolami}{Yes, indeed. According to my models, remnants with masses between 0.7 and 0.8~M$_{\odot}$ (the largest ones I computed) cross the HR diagram in only $\sim$~100 to $\sim$~10~yr and then start to decrease their luminosity towards the white dwarf phase. So they should be very rare. In addition, one might wonder whether they would not be still highly obscured by circumstellar material.}

\discuss{Ventura}{Which are the typical timesteps and mass-loss rates which you use during the transition time?}

\discuss{Miller Bertolami}{During the transition phase I forced the code not to use the extremely small timesteps our algorithm would naturally suggest (which might be as small as 10$^{-4}$~yr). So I usually forced timesteps to be between 0.1 and 1~yr during the transition phase.}

\end{discussion}

\end{document}